\documentclass[trackchanges,twocolumn]{aastex701} 

\usepackage{amssymb}
\usepackage{xcolor}
\newcommand{\cmark}{\textcolor{green!60!black}{$\checkmark$}}
\newcommand{\xmark}{\textcolor{red}{$\times$}}
\usepackage{fontspec}

\usepackage{comment}

\begin{document}

\title{Learning the Universe at High Redshifts: Impact of Accretion Modeling on Early Black Hole Growth}

\author[orcid=0000-0002-0275-3001]{Jonathan Kho}
\affiliation{University of Virginia 
530 McCormick Rd
Charlottesville, VA 22904, USA}
\affiliation{Virginia Institute for Theoretical Astronomy, University of Virginia, Charlottesville, VA 22904, USA}
\affiliation{The NSF-Simons AI Institute for Cosmic Origins, USA}
\email[show]{yja6qa@virginia.edu}  

\author[0000-0002-7080-2864]{Aklant K. Bhowmick}
\affiliation{University of Virginia 
530 McCormick Rd
Charlottesville, VA 22904, USA}
\affiliation{Virginia Institute for Theoretical Astronomy, University of Virginia, Charlottesville, VA 22904, USA}
\affiliation{The NSF-Simons AI Institute for Cosmic Origins, USA}
\email{whm4dg@virginia.edu}

\author[0000-0001-6260-9709]{Rainer Weinberger}
\affiliation{Leibniz Institute for Astrophysics Potsdam (AIP), An der Sternwarte 16, 14482 Potsdam, Germany}
\email{rweinberger@aip.de}

\author[0000-0002-5653-0786]{Paul Torrey}
\affiliation{University of Virginia 
530 McCormick Rd
Charlottesville, VA 22904, USA}
\affiliation{Virginia Institute for Theoretical Astronomy, University of Virginia, Charlottesville, VA 22904, USA}
\affiliation{The NSF-Simons AI Institute for Cosmic Origins, USA}
\email{ygn5rz@virginia.edu}

\author[0000-0002-2183-1087]{Laura Blecha}
\affiliation{Department of Astronomy, University of California Berkeley, Gainesville, FL 32611, USA}
\email{lblecha@ufl.edu}

\author[0000-0001-6950-1629]{Lars Hernquist}
\affiliation{Harvard-Smithsonian Center for Astrophysics, Harvard University, Cambridge, MA 02138, USA}
\email{lhernquist@cfa.harvard.edu}

\author[0000-0003-2630-9228]{Greg L. Bryan}
\affiliation{Department of Astronomy, Columbia University
1328 Pupin Physics Lab, MC 5246
550 West 120th Street
New York, NY 10027}
\email{gb2141@columbia.edu}

\author[0000-0002-8111-9884]{Alex M. Garcia}
\affiliation{University of Virginia 
530 McCormick Rd
Charlottesville, VA 22904, USA}
\affiliation{Virginia Institute for Theoretical Astronomy, University of Virginia, Charlottesville, VA 22904, USA}
\affiliation{The NSF-Simons AI Institute for Cosmic Origins, USA}
\email{aku7cf@virginia.edu}

\author[0009-0002-1233-2013]{Niusha Ahvazi}
\affiliation{University of Virginia 
530 McCormick Rd
Charlottesville, VA 22904, USA}
\affiliation{Virginia Institute for Theoretical Astronomy, University of Virginia, Charlottesville, VA 22904, USA}
\affiliation{The NSF-Simons AI Institute for Cosmic Origins, USA}
\email{dkk9en@virginia.edu}

\author[0000-0003-4546-3810]{Alejandro Saravia}
\affiliation{University of Virginia 
530 McCormick Rd
Charlottesville, VA 22904, USA}
\affiliation{National Radio Astronomy Observatory, 520 Edgemont Road, Charlottesville, VA 22903, USA}
\affiliation{Universidad Nacional Autónoma de Honduras, Edificio K1, Ciudad Universitaria, Tegucigalpa, Honduras}
\email{mas6um@virginia.edu}


\author[0000-0003-4597-6739]{Boon Kiat Oh}
\affiliation{School of Physics, Korea Institute for Advanced Study, 85 Hoegiro, Dongdaemun-gu, Seoul 02455, Republic of Korea}
\affiliation{Department of Physics, University of Connecticut, 196 Auditorium Road, U-3046, Storrs, CT 06269-3046, USA}
\email[]{bkoh@kias.re.kr}

\begin{abstract}

JWST discoveries of the earliest ($z \gtrsim 9$) supermassive black holes (BHs, $M_\bullet \gtrsim 10^6\,\rm{M}_\odot$) challenge the BH seeding and accretion models of most cosmological simulations. 
In this work, we compare early BH growth arising from three different accretion prescriptions characterized by distinct scalings between the accretion rate ($\dot{M}_{\rm \bullet}$) and the BH mass ($M_{\rm \bullet}$): the commonly used Bondi–Hoyle model ($\dot{M}_{\rm \bullet}\propto M_{\rm \bullet}^2$), and two free-fall models with shallower scalings ($\dot{M}_{\rm \bullet}\propto M_{\rm \bullet}^{1/2}$ and $M_{\rm \bullet}$).
Bondi accretion tends to produce stronger runaway growth than the free-fall models when using heavy ($\sim10^5\,\rm{M}_\odot$) seeds in extreme environments owing to the steeper $M_\bullet$ scaling, but its sensitivity to the local gas sound speed makes it more susceptible to suppression from temperature increases due to AGN and stellar feedback.
The free-fall models tend to produce stronger growth for lower-mass seeds ($\sim10^{3-4}\,\rm{M}_\odot$) in moderate environments as they are less dependent on the BH's mass to accrete effectively, however in this regime BH growth remains negligible for all accretion models in the presence of fiducial stellar feedback.
Enhancing early BH growth via many BH--BH mergers disproportionately enhances subsequent accretion-driven growth for Bondi due to the steeper $M_{\rm \bullet}$ dependence. 
Our simulations can thus assemble BHs with masses of $\sim10^6-10^7~M_{\odot}$ at $z\gtrsim9$, as inferred by JWST, under two circumstances: 1) abundant heavy-seed formation that drives BH--BH mergers, or 2) Bondi accretion with weak feedback.

\end{abstract}


\section{Introduction} 

Before JWST, the discovery of $\sim10^9~M_{\odot}$ BHs powering the brightest quasars at $z\sim6-7$ already posed significant challenges to models of early BH formation and growth~\citep{fan_survey_2001, fan_survey_2003, li_formation_2007, volonteri_formation_2012}. 
With the advent of JWST, these challenges have been further exacerbated as active galactic nuclei (AGN) are now being discovered at even higher redshifts~(up to $z\sim9-11$) with inferred masses in the range $10^6~{\rm M}_\odot < M_\bullet < 10^8 ~{\rm M}_\odot$~\citep{Larson_etal_2023, Maiolino_etal_2024, Natarajan_etal_2024, Bodgan_etal_2024, Kovacs_etal_2024}. 
In addition to pushing the distance frontier, JWST has also uncovered a new class of compact objects largely at $z\sim4-7$ with distinctive ``V-shaped'' spectral energy distributions, dubbed Little Red Dots~\citep[LRDs,][]{Matthee_etal_2024, Kocevski_etal_2025, Umeda_etal_2026, Madau_etal_2026}. 
Most spectroscopically confirmed LRDs exhibit broad emission lines that strongly suggest the presence of an AGN, while simultaneously showing striking differences from standard AGN populations, such as weak X-ray, far-IR, and radio emission~\citep{ananna___x-ray_2024,yue_stacking_2024, casey_dust_2024, xiao_no_2025, perger_deep_2025, setton_confirmed_2025}. 
Furthermore, several studies infer BH masses and BH-to-stellar mass ratios that are substantially higher than expectations from local BHs~\citep{pacucci_jwst_2023, maiolino_jades_2024, Juodzbalis_etal_2026}, while others argue that these BH masses may be systematically overestimated~\citep{Naidu_etal_2025, Rusakov_etal_2026}. 
These discoveries could pose stringent challenges to our understanding of early BH formation and growth.

Cosmological hydrodynamic simulations are among the most powerful theoretical tools for making predictions of BH~(and galaxy) populations to compare against observations~\citep{Dubois_etal_2012, Sijacki_etal_2015, McAlpine_etal_2016, Weinberger_etal_2018, Thomas_etal_2019}. 
However, owing to resolution limitations, many components of galaxy formation models can only be implemented via sub-grid prescriptions that carry substantial uncertainties associated with unresolved small-scale physics. 
Examples of such prescriptions include implementations of BH seeding, dynamics, gas accretion and AGN feedback, as well as star formation and stellar feedback, all of which have large impacts on the earliest BH growth.
In most cosmological simulations, these subgrid models are generally calibrated to reproduce BH and galaxy populations in the local Universe ($z\sim0$), and thus tend to show good agreement with observations at low redshift. 
However, results can diverge strongly between simulations at higher redshifts, as different approaches to BH seeding, accretion, and AGN feedback lead to substantially different BH growth histories and increasing discrepancies with observations \citep{Habouzit_etal_2021, Habouzit_etal_2022}.

Despite the diversity of galaxy formation models, a large fraction of cosmological simulations fail to produce sufficiently early BH growth to explain the high-redshift BH populations currently being uncovered by JWST~\citep[][see their Figure 4]{2026arXiv260500092I}. 
While these measurements may undergo substantial revisions in the future, they nevertheless provide strong motivation for a systematic exploration of the interplay between BH seeding, accretion, environment and feedback in determining where and how SMBHs can form and grow. 
For example, the recent \texttt{BRAHMA} simulations~\citep{Bhowmick_etal_2024b, Bhowmick_etal_2024c} systematically explored the impact of a broad range of physically motivated BH seeding models on early BH growth while leaving the other aspects of the galaxy formation model unchanged from its predecessor \texttt{IllustrisTNG}. 
\cite{Bhowmick_etal_2026} identified that one possible pathway to producing the currently inferred masses of some of the most distant $z\sim9-11$ JWST BHs involves a very lenient production of heavy $\sim10^5~M_{\odot}$ seeds which boosts early BH growth through BH--BH mergers. 
\cite{Bhowmick_etal_2025} and \cite{Kho_etal_2025} also found that this lenient production of heavy seeds can produce BH populations which have a high-$z$ $M_\bullet-M_*$ relation that is overmassive relative to the local relation and yet a high-$z$ $M_\bullet-\sigma$ relation consistent with the local relation; this has been tentatively suggested by JWST observations \citep{Juodzbalis_etal_2026}.
However, it remains unclear whether heavy seeds can form in such abundance, since one of the most popular candidate mechanisms---direct collapse black holes~\citep{bromm_formation_2003, begelman_formation_2006, lodato_supermassive_2006, lodato_mass_2007, luo_direct_2018}---is generally predicted to be relatively rare~\citep{dijkstra_fluctuations_2008,  shang_supermassive_2010,agarwal_ubiquitous_2012, dijkstra_feedback-regulated_2014, habouzit_number_2016, Bhowmick_etal_2022}. 
Nevertheless, there are some recently proposed channels that could make heavy seed production more efficient, such as stellar collisions in ultra-dense nuclear star clusters ~\citep{kritos_massive_2023,pacucci_2025}, or intermittent growth of a small fraction of Pop III stellar remnants~\citep{mehta_growth_2026}. 
In any case, it is also imperative to explore how other aspects of BH physics impact early BH growth and to possibly identify a broader range of pathways for explaining the JWST observations.

Aside from BH seeding, another important component of modeling BHs in cosmological simulations is gas accretion onto BHs. 
By far the most common implementation in most cosmological simulations is the Bondi-Hoyle model \citep{1944MNRAS.104..273B, 1952MNRAS.112..195B}. 
This approach is popular for its relatively simple implementation and its ability to be calibrated to reproduce local BH-galaxy scaling relations.
However, the Bondi accretion model is built on physical assumptions that likely do not reflect the conditions surrounding SMBHs, especially in the high-$z$ Universe.
In particular, it assumes radial inflow onto the accreting BH, a situation that would be immediately disrupted by the presence of a moderate amount of angular momentum in the surrounding gas.
It is also derived assuming gas capture occurs from a hot, diffuse uniform background, a situation which is unlikely to be realistic for high-$z$ galactic nuclei \citep{dekel_formation_2009}.
Furthermore, the Bondi model has a steep scaling between the accretion rate and BH mass,~i.e. $\dot{M_\bullet}\propto M_{\bullet}^2$, where $M_{\bullet}$ is the BH mass. 
As a result, it leads to a disproportionately stronger growth of higher mass BHs compared to lower mass BHs, which is not ideal for the earliest stages of BH seed growth. 
Bondi accretion also depends on the local sound speed, which often has a lower limit imposed by the commonly adopted effective equation of state of the ISM~\citep[e.g.,][]{Springel_&_Hernquist_2003}.
In light of these limitations, new alternatives to the Bondi model have been developed. 
One prominent alternative is the torque-limited accretion~(TLA) model.
This model presumes that non-axisymmetric perturbations in a galaxy's potential well drive gas inflow to the galactic nucleus, thereby enabling the central BH to grow via accretion.
This model has been motivated by e.g., \cite{hopkins_how_2010, Hopkins_Quataert_2011}, who use multi-scale galaxy simulations to track the inflow of gas down to the central BH and find that a TLA analytic/subgrid model more faithfully captures the true gas inflow rates in their simulations than Bondi.
\cite{Angles-Alcazar_2015} also employ a TLA model in their simulations.
They find that their model (which scales as $ \dot{M_\bullet}\propto M_\bullet^{1/6}$) naturally produces galaxies that converge to the $\rm M_\bullet-M_{bulge}$ relation without invoking mass averaging via mergers or regulatory feedback processes. 

However, one major difficulty in implementing the TLA model is that it requires the ability to perform a robust bulge-disk decomposition of galaxies, and thereby estimate the degree of non axisymmetric perturbations that generate the gravitational torques. 
This is particularly challenging at high redshift, where disks and bulges are often not well formed. 
Additionally, these non axisymmetric perturbations may themselves not be fully resolved in cosmological simulations. 

\cite{Weinberger2025} recently developed a new flexible free-fall accretion model that does not attempt a bulge-disk decomposition, and instead adopts tunable parameters intended to account for both resolved and unresolved non axisymmetric perturbations in the galaxy's gravitational potential. 
Furthermore, the model is parameterized to absorb the effects of other unresolved physical mechanisms driving mass inflow and accretion, thereby providing a simple and tunable prescription informed by smaller-scale simulations.
One of the key parameters is an adjustable power-law dependence of the BH accretion rate on BH mass~($M_{\bullet}^{\rm n}$). 
\cite{Weinberger2025} presented two variants of the free-fall model that successfully reproduce the observed local BH number densities, corresponding to power-law dependencies substantially shallower than Bondi accretion~($\rm n = 0.5,1$).

In this work, we present a new suite of simulations dedicated to exploring high-redshift BH growth by combining the \texttt{BRAHMA} seeding models with the novel free-fall accretion models. 
In particular, we use the two calibrated variants of the free-fall model presented in \cite{Weinberger2025}. 
We compare these results against the traditional Bondi accretion model adopted in the original \texttt{BRAHMA} simulations, and examine the implications for the recent JWST AGN discoveries as well as the brightest $z\sim6$ quasars discovered prior to JWST. 
We compare the different accretion models across a wide range of environments, including typical regions of the high-$z$ Universe as well as the most extreme overdense peaks where the brightest $z\sim6$ quasars are expected to reside. 
Finally, we explore variations in AGN and stellar feedback to understand their role in determining the relative performance of the different accretion models across different environments.


The rest of this paper is organized as follows. In section \ref{sec:Methods}, we describe our methods including initial conditions generation, our underlying galaxy formation model, and the different accretion models used.
In section \ref{sec:Results} we present the results of our simulations, and in section \ref{sec:Discussion} we discuss the implications of these results for early SMBH growth and observed JWST AGN.
We provide concluding remarks in section \ref{sec:Conclusions}.

\section{Methods}\label{sec:Methods}

\subsection{Initial conditions}

Our simulations were run using the {\sc arepo} gravity $+$ magnetohydrodynamic (MHD) solver \citep{2010MNRAS.401..791S, 2011MNRAS.418.1392P, 2016MNRAS.462.2603P, 2020ApJS..248...32W}. 
Gravity is calculated via a  particle mesh (PM) tree gravity solver \citep{1986Natur.324..446B}, and the evolution of the gas is described by the ideal MHD equations solved over a dynamic unstructured grid generated via a Voronoi tessellation of the domain. 
All of our simulations use a Planck Collaboration XIII (\citeyear{2016A&A...594A..13P}) cosmology i.e. $\Omega_\Lambda$ = 0.6911, $\Omega_\mathrm{m}$ = 0.3089, $\Omega_\mathrm{b}$ = 0.0486, $H_0 = 67.76\ \mathrm{km\ s^{-1}\ Mpc^{-1}}$, $\sigma_8$ = 0.8159, $n_s$ = 0.9667. 
We identify halos using the Friends-of-Friends (FOF) algorithm \citep{1985ApJ...292..371D} with a linking length of 0.2 times the mean particle separation, and subhalos are identified using the \texttt{SUBFIND} \citep{2001MNRAS.328..726S} algorithm. 

We use three different initial conditions (ICs) for our simulations targeting three different environments: 

\begin{itemize}

\item The ``Large Halo'' IC is a $9~\rm Mpc/h$ box constrained to be a rare overdense region that forms a compact halo with mass $\sim 3\times10^{12}~\rm{M}_\odot$ by $z=6$. These ICs have a target baryon mass resolution of $2.5\times 10^5~\rm{M}_\odot$, and we seed BHs with an initial mass of $M_{\rm seed} = 10^5~\rm{M}_\odot$. 
This constrained IC is generated using the \texttt{GaussianCR} package~\citep{Ni2022}.

\item The ``Medium Halo'' IC features a zoom-in region which forms a $4\times10^{11}~\rm{M}_\odot$ halo by $z=6$, with a baryon mass resolution of $2.9\times10^4~\rm{M}_\odot$. 
In these ICs, we adopt a BH seed mass of $M_{\rm seed} = 1.25\times10^4~\rm{M}_\odot$.
These ICs were generated using the MUSIC (MUlti-Scale Initial Conditions) program \citep{hahn_multi-scale_2011}.

\item The ``Small Halo'' IC is a random realization within a $3.125~\rm Mpc/h$ box, representing a typical, unconstrained small region of the Universe. 
The most massive halo formed in this box by $z=6$ has a mass of $\sim 1.5\times10^{10}~\rm{M}_\odot$. 
The target baryon mass resolution is $3.64\times10^3~\rm{M}_\odot$, and we seed BHs with a mass of $M_{\rm seed} = 1.5625\times10^3~\rm{M}_\odot$.
These ICs were also generated using MUSIC.

\end{itemize}

\begin{figure*}
    \centering
    \includegraphics[width=1.0\linewidth]{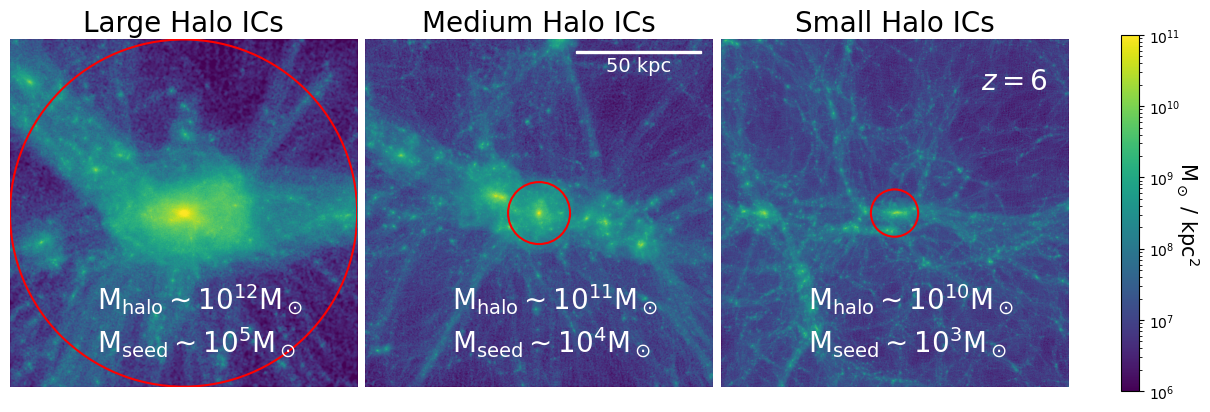}
    \caption{Distributions of DM surface densities in a cut through the most massive halo for the Large, Medium, and Small halo ICs each at $z=6$. We overplot in a red circle the virial radius of each of these target halos using the same physical scale, highlighting the extreme differences in the halos and halo environments between ICs.}
    \label{fig:Halo_dist}
\end{figure*}

Fig. \ref{fig:Halo_dist} visualizes the most massive halo in each of these ICs at $z=6$. 
The choice of these ICs, their resolutions, and the seed masses is motivated by both numerical and physical considerations. 
From a numerical standpoint, we decrease the mass resolution with increasing target $z=6$ halo mass to minimize our computational expense. 
This allows us to run enough simulations to study the impact of variations in seeding, feedback, and accretion. 
The choice of BH seed mass is primarily limited by the mass resolution of the simulations, as including BHs with masses much smaller than the surrounding gas and star particles may result in BHs receiving spuriously large gravitational kicks. 
Therefore, the ``Large Halo'' ICs only focus on the heaviest $\sim10^5~M_{\odot}$ seeds, whereas the ``Medium Halo'' and ``Small Halo'' ICs focus on smaller seeds. 

\begin{figure}
    \centering
    \includegraphics[width=0.95\linewidth]{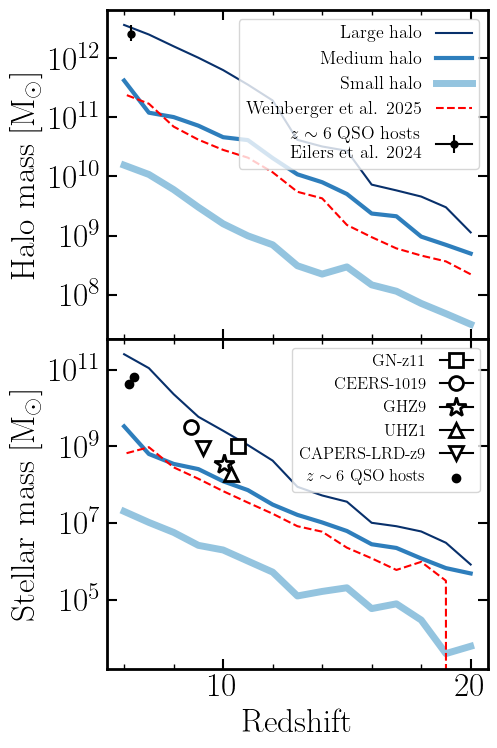}
    \caption{Halo and stellar mass evolutions for the most massive halo in each of our three ICs. 
    By $z=6$, the most massive halo in the large, medium, and small halo ICs has reached a total mass of $\sim3\times10^{12}, 4\times10^{11},$ and $1.5\times10^{10} \rm{M}_\odot$ respectively and a stellar masses of $\sim2\times10^{11}, 3\times10^{9},$ and $2\times10^{7} \rm{M}_\odot$ respectively. 
    Our Large Halo ICs are chosen to target the expected environments of the brightest $z\sim6$ quasars, but they also broadly reproduce the stellar masses of some of the JWST AGN we compare to, particularly GN-z11 and CEERS-1019. 
    Therefore, we shall compare the BH mass evolution produced in this IC to observed BH mass measurements for both $z\sim6$ quasars \citep{eilers_eiger_2024} and $z\sim9-11$ JWST AGN.}
    \label{fig:mostmassivehalo}
\end{figure}

In Fig.~\ref{fig:mostmassivehalo}, we show the mass growth histories of the most massive halo in each of these ICs. 
Our Medium Halo ICs target a halo with an assembly history most similar to the most massive halo in the uniform $12.5~\rm Mpc/h$ box from \citealt{Weinberger2025} (red dashed line). 
However, as we can clearly see, the $z\sim9$--11 galaxies produced by the Medium Halo ICs are less massive than current estimates from JWST. 
Only the rare, extreme environment probed by the Large Halo IC produces $z\sim9$--11 galaxies broadly consistent with current JWST measurements. 
Finally, our ``Small Halo'' ICs complement the other two ICs such that, taken together, we are able not only to understand the implications for the JWST measurements, but also to probe the interplay between accretion scalings, feedback processes, and merger-driven growth across a wider range of environments than was possible in \cite{Weinberger2025}.

\subsection{Galaxy Formation Model}

With the exception of BH seeding, accretion, and dynamics, we adopt the galaxy formation model used in the \texttt{IllustrisTNG} (hereafter \texttt{TNG}) simulations~\citep{Weinberger_etal_2017, Pillepich_etal_2018a}, which itself is based on the \texttt{Illustris} model~\citep{Vogelsberger_etal_2013, torrey_model_2014}.
Radiative cooling is calculated for contributions from primordial species ($\mathrm{H, H^+, He, He^+, He^{++}}$) as per \cite{Katz_etal_1996} and metals as per \cite{Wiersma_etal_2009}, which are calculated in the presence of a spatially uniform, time-dependent UV background and interpolated from tables. 
Gas cells above a star formation threshold density are made to represent a pressurized, multiphase interstellar medium via an effective equation of state \citep{Springel_&_Hernquist_2003}.
Star formation occurs as gas cools and condenses to exceed a density of $0.13\ \mathrm{cm}^{-3}$.
Star particles represent stellar populations with a single age and metallicity, and have an initial stellar mass function taken from \cite{Chabrier_2003}.
Subsequent stellar evolution is based on \cite{Vogelsberger_etal_2013} with modifications described in \cite{Pillepich_etal_2018a}.
Stellar chemical enrichment follows the evolution of H, He, C, N, O, Ne, Mg, Si, and Fe, and supernova feedback is modeled as a galactic-scale wind that deposits metals, mass, and momentum to nearby gas cells.

\subsubsection{Black Hole Seeding and Dynamics}

We use the BH seeding prescriptions developed by \cite{Bhowmick_etal_2021, Bhowmick_etal_2022, Bhowmick_etal_2024a}, which are designed to emulate a wide range of seed formation scenarios.
In particular, these scenarios enforce seeds to form within high-density and low-metallicity gas at early times. 
Seeds are generally formed in halos above a threshold in total mass~($M_h$) as well as star forming~\&~metal-poor gas mass~($M_{\rm sfmp}$). 
In this work, we primarily focus on two broad categories of seed models: a strict model and a lenient model.

The strict model adopts seeding thresholds of $M_h = 1000~M_{\rm seed}$~\&~$M_{\rm sfmp} = 1000~M_{\rm seed}$ for all the ICs. 
The goal is to restrict seed formation such that no or very few BH--BH mergers occur, enabling us to understand the impact of accretion and feedback processes on the growth of a single seed.

The lenient model adopts smaller seeding thresholds compared to the strict models. For the Small Halo and Medium Halo ICs, we reduce the dense \& metal-poor gas mass threshold, i.e. $M_h = 1000~M_{\rm seed}$~\&~$M_{\rm sfmp} = 5~M_{\rm seed}$. 
For the Large Halo ICs, we use $M_{\rm sfmp} = 5~M_{\rm seed}$, while fully removing the explicit halo mass threshold. 
This is because the lower mass resolution of the simulation already imposes an implicit halo mass seeding threshold ($M_h \geq 6.33 \times 10^7 \rm{M}_\odot$ corresponding to the minimum of 32 DM particles needed for FOF halo identification). 
In these lenient models, we expect an enhanced contribution to BH growth from BH--BH mergers, and we study how this also influences gas accretion under different accretion models.

As used in \cite{Bhowmick_etal_2026}, we employ the subgrid dynamical friction model developed by \cite{Ma_etal_2023} which compensates for missing dynamical friction forces as a consequence of poor mass and spatial resolution. 
This is done by adding a corrective factor to each BH particle's acceleration for each mass element encountered in the normal gravity PM tree calculation. 
The form of this corrective factor is as follows:

\begin{equation}
    \mathbf{a}_{\mathrm{df} } = \sum_i \frac{\alpha_i b_i}{(1 + \alpha_i^2)(r_i + r_{\mathrm{soft}})}\left(\frac{G\Delta m_i}{(r_i+r_{\mathrm{soft}})^2}\right)\hat{\mathbf{V}}_i,
\end{equation}
where $\alpha_i\approx b_iV_i^2/GM_\bullet$ and $b_i \equiv r_i|\bm{\hat{r}}_i - (\bm{\hat{r}}_i\cdot\hat{\bm{V}}_i)\hat{\bm{V}}_i|$, and $\Delta m_i$, $r_i$, and $V_i$ are the mass, relative distance, and relative velocity of the $i^\mathrm{th}$ mass element in the gravity tree.
Two BHs are merged if they are considered to be gravitationally bound to each other and are separated by less than two times the BH softening length (0.5, 0.25, and 0.125 ckpc for the Large, Medium, and Small ICs respectively). 
The criterion for two BHs to be gravitationally bound to each other is $|\Delta v|^2/2 + \Delta a\Delta r < 0$, where $\Delta v$, $\Delta a$, and $\Delta r$ are the BHs' relative velocities, accelerations, and distances respectively.

\subsubsection{Black Hole Accretion}\label{sec:accretion_models}

In this work, we investigate early BH growth under three different accretion models with distinct dependences of the BH accretion rate~($\dot{M}_{\bullet}$) on the BH mass~($M_{\bullet}$) as well as the surrounding gas properties. 
These include the most commonly adopted Bondi-Hoyle accretion model~\citep{1944MNRAS.104..273B,1952MNRAS.112..195B}, and two variations of a free-fall-based accretion model~\citep{Weinberger2025}.

Bondi-Hoyle accretion is given by
\begin{equation}
    \dot{M}_{\mathrm{Bondi}} = \alpha_{\rm{boost}}\frac{~4\pi G^2 M_{\bullet}^2 \rho}{c_s^3}
\end{equation}
where $G$ is the gravitational constant, and $\rho$ and $c_s$ are the gas density and sound speed averaged over a kernel comprising 256 nearest gas neighbors. 
We adopt a boost factor of $\alpha_{\rm{boost}}=100$ to account for unresolved multiphase gas as commonly adopted in previous works \citep{Booth_Schaye_2009}. 
This estimate assumes spherically symmetric accretion onto the BH and the accretion rate is ultimately set by the competition between the BH's gravity and the thermal pressure of the gas.

The free-fall accretion model assumes that the BH accretion rate is limited by the angular momentum of the cold gas phase, and that angular-momentum loss driven by non-axisymmetric perturbations in the gravitational potential leads to subsequent gas infall~\citep{Weinberger2025}. 
The key assumption is that this angular-momentum transport is so efficient that a fraction of the gas can lose enough angular momentum to fall onto the BH within a free-fall time set by the local gravitational potential at the gas location.
More precisely, for every gas cell $i$ with mass $m_i$, we estimate its contribution to the accretion rate $\dot{m}_i$ as
\begin{equation}
    \dot{m}_i = \eta \frac{m_i}{t_{\rm ff}}, \\
    t_{\rm ff} = \left(\frac{d^3}{G M_{\bullet}}\right)^{1/2},
\end{equation}
where the free-fall time $t_{\rm ff}$ is computed from a fixed distance $d$ to the BH. 
The efficiency parameter $\eta$ captures the strength of non-axisymmetric perturbations as well as a possible radial dependence of the inflow rate, and is given by
\begin{equation}
    \eta = A\left(\frac{d}{R_s}\right)^\alpha,
\end{equation}
where $R_s = {2GM_{\rm BH}}/{c^2}$ is the Schwarzschild radius. 
The accreted mass from gas cell $i$ over a timestep from $t_n$ to $t_{n+1}$ is then
\begin{equation}
    \Delta m_i = \int_{t_n}^{t_{n+1}} \dot{m}_i \, dt,
\end{equation}
which is summed over all gas cells to obtain the total accreted mass $\Delta m = \sum_i \Delta m_i$.

The model is described by two free parameters, $A$ and $\alpha$. 
$A$ sets the normalization of the inflow rate on resolved scales, while $\alpha$ controls the radial dependence of the inflow down to the Schwarzschild radius. 
Different values of $\alpha$ produce different scalings between BH mass and the accretion rate.

\cite{Weinberger2025} identified two variations of the free-fall model that reproduce the observed $z=0$ BH mass density (BHMD). 
The first is the ``ff'' (freefall) model characterized by $A = 0.001$ and $\alpha = 0$, corresponding to relatively low inflow rates on large scales but no suppression on small scales, yielding $\dot{M}_{\bullet} \propto \rho~M_{\bullet}^{1/2}$. 
The second is the ``modff'' (modified freefall) model with $A = 100$ and $\alpha = -0.5$, which assumes higher inflow rates on large scales but significant suppression on small scales, resulting in $\dot{M}_{\bullet} \propto \rho~M_{\bullet}$.

To summarize, we compare the impact of three accretion models (Bondi, ff, and modff) on early BH growth. 
These models exhibit distinct dependencies on BH mass ($M_{\bullet}^2$ vs.\ $M_{\bullet}$ vs.\ $M_{\bullet}^{1/2}$) as well as on their environmental properties. 
While the Bondi accretion rate depends on both density and sound speed, the free-fall models depend primarily on the gas density. 
We explore how these differences affect the interplay between accretion, feedback, and BH seeding.
In all of our simulations we limit the accretion rates by the Eddington limit.

\subsubsection{AGN and Stellar Feedback}

To model the feedback produced by AGN, accreting BHs are assumed to radiate at bolometric luminosities $L_{\rm Bol} = \epsilon_r \dot{M}_{\bullet} c^2$, where $\epsilon_r=0.1$ is the radiative efficiency, and a fraction of the energy released is assumed to couple to the ISM as feedback. 
This coupling is implemented in the same way as in the \texttt{TNG} model, whereby BHs accreting at a ``high'' Eddington ratio emit thermal feedback by depositing a fraction of their radiated luminosity to nearby gas cells, and BHs accreting at ``low'' Eddington ratios emit kinetic feedback by injecting kinetic energy at irregular intervals into nearby gas cells \citep{Weinberger_etal_2017}. 
The Eddington ratio cutoff for the thermal vs. kinetic feedback modes is given by

\begin{equation}\label{eqn:AGN_cutoff}
    \chi = \mathrm{min}\left[0.002\left(\frac{M_{\bullet}}{10^8\rm{M}_\odot}\right)^2,0.1\right],
\end{equation}
and the energy injected into the surroundings is given by $\dot{E}_{\mathrm{therm}} = 0.02 \dot{M}c^2$ for thermal feedback and $\dot{E}_{\mathrm{kin}} = \epsilon_{\mathrm{kin}}\dot{M}c^2$, where 
\begin{equation}
    \epsilon_{\mathrm{kin}} = \min\left(\frac{\rho}{\rho_{\mathrm{SF}}},0.2\right),
\end{equation}
and ${\rho_{\mathrm{SF}}}$ is the density required for the onset of star formation. In practice, the kinetic feedback becomes important only for massive $\gtrsim10^8~M_{\odot}$ BHs at low-redshift. 
In this work which focuses on early BH growth, it is primarily the thermal feedback that plays a dominant role in regulating BH accretion.
We explore the impact of turning this thermal AGN feedback off under different accretion models in Section~\ref{sec:Results}.

BH growth can also be impacted by stellar feedback, which is implemented as galactic-scale winds released isotropically by star-forming gas cells \citep{Pillepich_etal_2018a}.
These winds hydrodynamically decouple from the surrounding gas and recouple once the ambient gas has become sufficiently diffuse or enough time has passed.
The ``mass loading factor" for these winds is defined to be
\begin{equation}
    \eta = \frac{\dot{M}_{\mathrm{wind}}}{\dot{M}_{\mathrm{SFR}}} = \frac{2}{v_w^2}e_w(1 - \tau_w),
\end{equation}
where $\dot{M}_{\mathrm{SFR}}$ is the local star formation rate, $\dot{M}_{\mathrm{wind}}$ is the gas mass to be converted into winds, $v_w$ is the wind velocity, $\tau_w$ is the fraction of wind energy that is thermal (as opposed to kinetic), and $e_w$ is the wind energy available to a star-forming gas cell.
This wind energy $e_w$ depends on the gas metallicity ($Z$), the number of type II Supernovae (SNe) per formed stellar mass ($N_{\rm SNII}$), and the energy released per SN ($E_{\rm SNII}$) \citep[see eqn. 3 of][]{Pillepich_etal_2018a}.
In nearly half of our simulations we test the impact of reducing stellar feedback on BH growth by reducing the SNe energy available from $3.6\times 10^{51}$ erg (the fiducial \texttt{TNG} value) to $0.36 \times 10^{51}$ erg (10\% of the fiducial \texttt{TNG} value).
The scripts used for the analysis in this work are publicly available via Zenodo \citep{kho_2026_20561794}.

\subsection{Simulation suite}

In total, we run 45 simulations down to $z=6$ spanning 3 ICs targeting different mass halos, 3 distinct accretion models (Bondi, ff, and modff), 4 different feedback variations (AGN \& stellar, AGN \& low stellar, no AGN \& stellar, and no AGN \& low stellar), and 2 seeding prescriptions (lenient vs. strict).
This allows us to test the performance of each of the 3 accretion models across a wide range of seed and halo masses, feedback scenarios, and seeding prescriptions.
We provide a summary of all of the simulations presented in this work in Table \ref{tab:Simulation_summary}.

\begin{table}[h!]
\centering

\begin{tabular*}{\columnwidth}{@{\extracolsep{\fill}}|c|c|c|c|}

\hline
\multicolumn{4}{|c|}{\textbf{Restrictive Seeding}} \\
\hline
\multicolumn{4}{|c|}{Accretion models: ff, modff, Bondi} \\
\hline
ICs & Seed Mass [$\rm{M}_\odot$] & AGN & Energy per SN \\
 & & & [$10^{51}$ erg] \\
\hline

       &        & \cmark & 3.6 \\
Large  & $10^5$ & \cmark & 0.36 \\
       &        & \xmark & 3.6 \\
       &        & \xmark & 0.36 \\
\hline

       &        & \cmark & 3.6 \\
Medium & $1.25\times10^4$ & \cmark & 0.36 \\
       &        & \xmark & 3.6 \\
       &        & \xmark & 0.36 \\
\hline

       &        & \cmark & 3.6 \\
Small  & $1.5625\times10^3$ & \cmark & 0.36 \\
       &        & \xmark & 3.6 \\
       &        & \xmark & 0.36 \\
\hline
\end{tabular*}

\vspace{0.3cm}

\begin{tabular*}{\columnwidth}{@{\extracolsep{\fill}}|c|c|c|c|}
\hline
\multicolumn{4}{|c|}{\textbf{Lenient Seeding}} \\
\hline
\multicolumn{4}{|c|}{Accretion models: ff, modff, Bondi} \\
\hline
ICs & Seed Mass [$\rm{M}_\odot$] & AGN & Stellar [$10^{51}$ erg]\\
\hline

Large  & $10^5$ & \cmark & 3.6 \\
Medium & $1.25\times10^4$ & \cmark & 3.6 \\
Small  & $1.5625\times10^3$ & \cmark & 3.6 \\

\hline
\end{tabular*}

\vspace{0.3cm}

\caption{Summary of seeding models and feedback configurations for our simulations. 
For each IC/seed mass and AGN and stellar feedback arrangement, we run 3 simulations for each of the 3 accretion models tested in this paper, e.g. for the Large Halo ICs with $10^5\rm{M}_\odot$ BH seeds, AGN feedback, and $3.6\times10^{51}$ erg per SN, we run 3 simulations using each of Bondi, ff, and modff for BH accretion. }
\label{tab:Simulation_summary}

\end{table}

\section{Results}\label{sec:Results}

We now compare early BH growth produced by our three different accretion models. 
In Section \ref{strict_sec}, we look at accretion-driven growth of individual seeds in the absence of BH–BH mergers, under different assumptions of AGN and stellar feedback.
For our Large Halo ICs with $\sim10^5~\rm{M}_\odot$ BH seeds, we also compare these BH growth histories to recent JWST measurements of BH masses at $z\sim10$ \citep{Larson_etal_2023, Maiolino_etal_2024, napolitano_dual_2025, Natarajan_etal_2024, taylor_capers-lrd-z9_2025}.
In Section \ref{sec:lenient}, we examine the role of abundant seeds and their mergers in shaping early BH growth across the different accretion models.

\subsection{Accretion-Driven BH Growth with Strict BH Seeding}
\label{strict_sec}

In Figures~\ref{fig:Constrained_BH} to \ref{fig:Small_BH}, we show the evolution of the mass~(left) and luminosity~(right) of the most massive BH in the simulations with strict seed models, such that the growth is driven purely by accretion. 
To illustrate the impact of different assumptions for AGN and stellar feedback, we divide the plots into $2\times2$ panels. The right columns correspond to lower stellar feedback, while the bottom rows correspond to no AGN feedback.

\subsubsection{Individual $\sim10^5~M_{\odot}$ seeds in Large Halo ICs}\label{sec:Large}

\begin{figure*}
    \centering
    \includegraphics[width=1.0\linewidth]{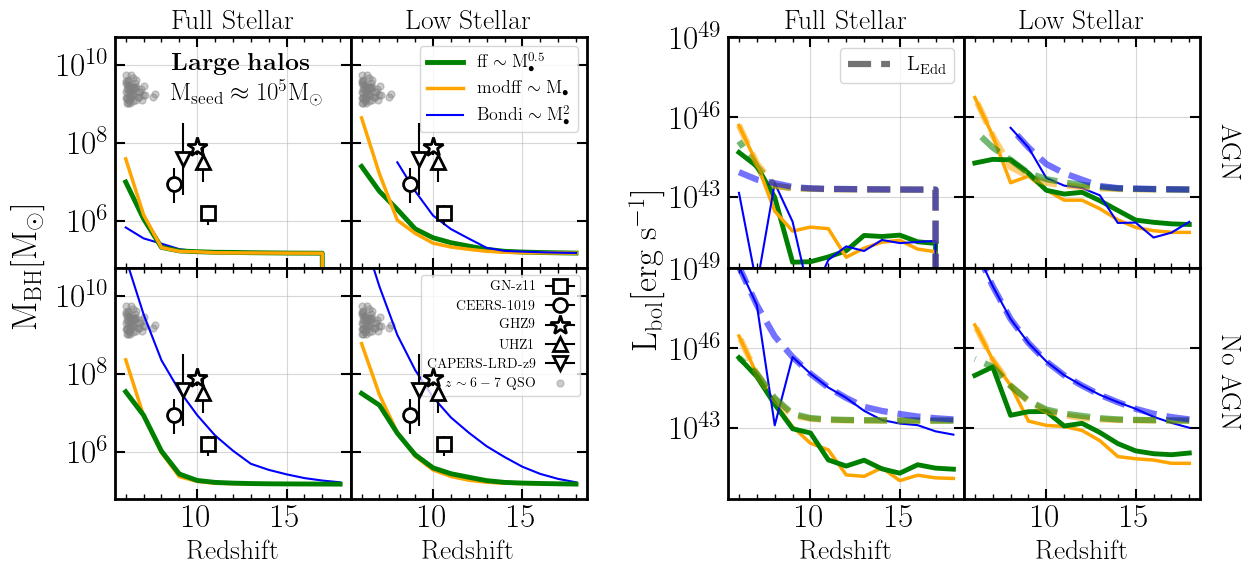}
    \caption{BH mass growth history for the most massive BH in each of the feedback and accretion model configurations for the large halo IC with the strict seed model. 
    For each subgroup of panels (left and right), the left columns correspond to simulations run with the fiducial stellar feedback strength, whereas the right columns were run with 10\% of the fiducial stellar feedback strength. 
    Across both subgroups of panels, the top rows correspond to runs with fiducial AGN feedback, and the bottom rows have AGN feedback turned off.
    When AGN or stellar feedback is reduced, growth via Bondi accretion is enhanced more than growth via ff or modff, enabling Eddington-limited growth by $z\sim13,15$ already for the no AGN, Bondi scenarios.
    This is due to stronger runaway BH growth from the steepest $M_{\bullet}^2$ scaling with BH mass.
    Under both fiducial AGN and stellar feedback, however, the free-fall models produce stronger BH growth than Bondi.
    This is because the inclusion of feedback tends to disproportionately suppress Bondi accretion compared to free-fall, since the former has an explicit dependence on sound speed.}
    \label{fig:Constrained_BH}
\end{figure*}

We first focus on the growth of a $10^5~M_{\odot}$ seed in the most massive halo in our Large Halo ICs~(Fig. \ref{fig:Constrained_BH}).
In the presence of fiducial AGN and stellar feedback~(top left panel of the mass evolutions), we generally find that none of the models produce any significant BH growth at $z\gtrsim9$. 
The BH growth is far below Eddington~($\lesssim0.01$) at this earliest stage~(top left panel in the luminosity evolution in Fig. \ref{fig:Constrained_BH}). 
While this has been previously reported by the \texttt{BRAHMA} simulations under the Bondi accretion model~\citep{Bhowmick_etal_2026}, here we show that this inefficiency in early accretion is also true for the free-fall models, despite having a much shallower scaling between accretion rate and BH mass.
At $z\lesssim9$, the BH masses start to show significant growth beyond the initial seed mass. 
In this regime, we find that both the ff and modff accretion models grow the BH at the Eddington rate, stronger than the Bondi accretion model that continues to be substantially sub-Eddington. 
The resulting final BH masses are $9.84\times 10^6\rm{M}_\odot$ and $3.87\times 10^7\rm{M}_\odot$ at $z=6$ for ff and modff models respectively, as opposed to $6.75\times10^5\rm{M}_\odot$ for Bondi.

When we look at the remaining panels of Fig. \ref{fig:Constrained_BH}, it becomes clear that the inefficiency of BH accretion at $z\gtrsim9$ is due to both stellar and AGN feedback.  
Interestingly, while feedback impacts the accretion for both Bondi and free-fall models, the impact on Bondi is disproportionately strong.
When we reduce stellar feedback (top right panel) or remove AGN feedback (bottom left panel), the Bondi accretion model begins generating substantial BH growth, overshooting the free-fall models.\footnote{We note that the Bondi simulation with fiducial AGN feedback and reduced stellar feedback failed to reach $z=6$, as it crashed due to unphysically high gas densities.}
This is primarily due to Bondi accretion’s stronger tendency toward runaway BH growth, driven by the steep $\dot{M}_{\rm \bullet} \propto M_{\rm \bullet}^2$ scaling. 
The free-fall models also show enhanced growth and an earlier onset of accretion, but the effect is much less pronounced because of their shallower BH mass scalings. 
Even among the free-fall models, the modff case produces stronger BH growth than the ff model because its accretion rate scales as $M_{\rm \bullet}$ rather than $M_{\rm \bullet}^{1/2}$.

A question now naturally emerges: if the Bondi model has a much stronger tendency for runaway BH growth compared to the free-fall models, why is it overtaken by the free-fall models when \textit{both} fiducial stellar and AGN feedback are applied~(upper left panels of Fig. \ref{fig:Constrained_BH})? 
We also note that this overtaking occurs at $z\lesssim7$, whereas at higher redshifts Bondi is still very slightly ahead of the free-fall models. 
As we show in Section \ref{sec: density and temperature}, this is a direct consequence of another fundamental difference between the Bondi and free-fall models, namely that the former has an inverse dependence on sound speed. 
Both AGN and stellar feedback lead to a sharp increase in gas temperature, which suppresses Bondi accretion but has no direct impact on free-fall accretion. 
When both of these feedback channels are active, the suppression of Bondi accretion is strong enough that it is overtaken by the free-fall models at $z\lesssim7$.

Finally, comparing the effects of removing AGN feedback versus reducing stellar feedback, we find that the free-fall models show similar levels of enhancement in both cases. 
However, for the Bondi model, we see a stronger enhancement when AGN feedback is removed compared to when stellar feedback is reduced. 
In fact, only for the Bondi model without AGN feedback do we see an immediate onset of Eddington accretion right after seed formation, which is largely sustained all the way to $z=6$~(bottom rows of Fig. \ref{fig:Constrained_BH}). 
For all other combinations of accretion and feedback scenarios, the earliest accretion remains well below Eddington.
As we show in the next subsections, this stronger sensitivity of Bondi accretion to AGN feedback, relative to stellar feedback, is unique to the growth of $\sim10^5~M_{\odot}$ seeds in the Large Halo ICs. 
As we go to smaller halos, stellar feedback tends to have a stronger impact on BH growth.

\subsubsection{Can Accretion onto Individual Seeds Explain the Observed High-$z$ BH Populations in Extreme, Rare Environments?}

As the Large Halo ICs correspond to extreme, rare environments that are likely hosts of the pre-JWST $z\sim6$ quasars, while also producing $z\sim9\text{--}11$ galaxies consistent with JWST, it is instructive to compare our predictions against these observations. 
In the presence of fiducial AGN and stellar feedback, BH growth is so inefficient that neither the $z\sim9\text{--}11$ JWST BH masses nor the $z\sim6$ quasar masses estimates are reproduced by any of the accretion models (see green, yellow, and blue lines in top left of left panel in Fig. \ref{fig:Constrained_BH}). 

Even when stellar and/or AGN feedback is reduced, none of the free-fall models can reproduce the JWST mass estimates. 
Only the modff model produces a $z\sim6$ BH that approaches the typical inferred mass of $z\sim6$ quasars, i.e. $\sim10^9~M_{\odot}$. 
The Bondi model has a much easier time producing the JWST BHs under reduced feedback. 
However, as shown in \cite{Bhowmick_etal_2026}, reducing stellar feedback also enhances the stellar masses of the host galaxies beyond the current JWST estimates, whereas reducing AGN feedback does not.

Overall, we find that among our Large Halo simulations that follow the accretion-driven growth of $\sim10^5~M_{\odot}$ seeds without BH--BH mergers, the current JWST BH mass estimates at $z\sim9-11$ are most readily reproduced under Bondi accretion with reduced AGN feedback. 
However, reducing AGN feedback is expected to lead to excessive BH growth and potentially unquenched star formation in massive galaxies at later times, since the model is calibrated to reproduce low-$z$ BHs and galaxies. 
Indeed, even by $z\sim6$, our ``no AGN feedback'' runs produce BHs with masses $\gtrsim10^{10}~M_{\odot}$, far exceeding those of observed $z\sim6$ quasars. 
Therefore, if the current JWST mass estimates are confirmed, and this pathway is responsible for their assembly, it becomes necessary to understand why AGN feedback coupling would be much weaker at early times, and how it subsequently strengthens to regulate BH growth and match the later BH populations. 
We speculate on potential physical mechanisms for this in section \ref{sec:implications}, but reserve an in-depth study for future work.

\subsubsection{Individual $\sim10^4~M_{\odot}$ seeds in Medium Halo ICs}

\begin{figure*}
    \centering
    \includegraphics[width=1.0\linewidth]{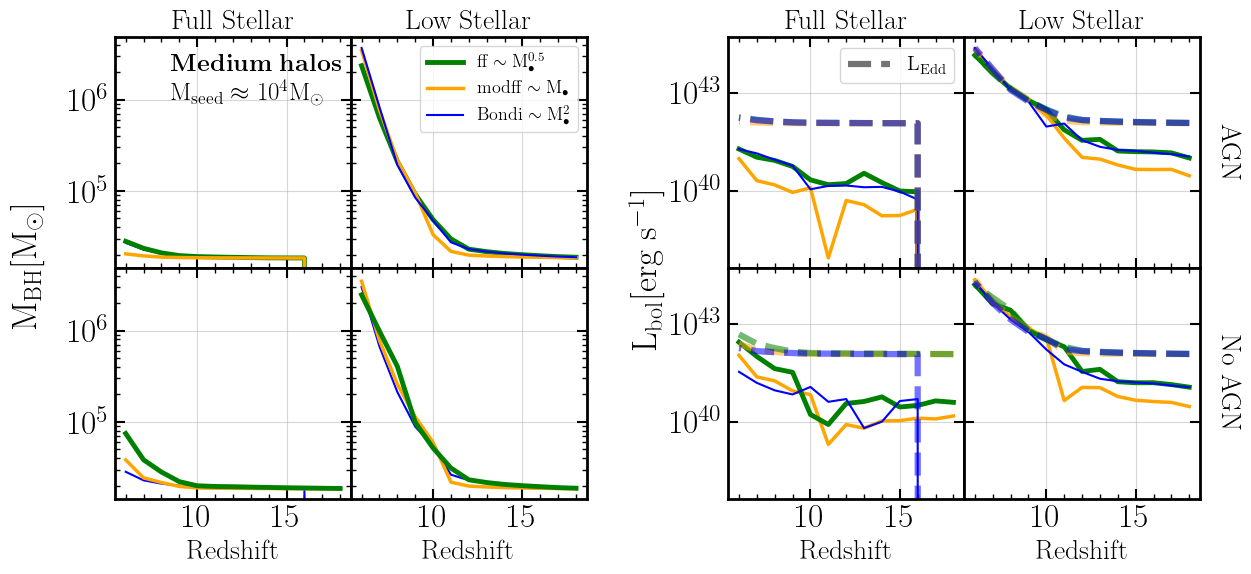}
    \caption{BH growth history for the most massive BH in our medium halo IC simulations with the strict seed model, formatted the same as in Fig. \ref{fig:Constrained_BH} with BH growth histories shown in the left panels and bolometric luminosities on the right for different combinations of AGN and stellar feedback.
    In these halos, removing AGN feedback has a minimal impact on all three accretion models, but reducing stellar feedback allows all three models to achieve Eddington-limited growth by $z=10$ regardless of AGN feedback.
    We do not include the JWST AGN shown in Fig. \ref{fig:Constrained_BH} here, as none of the BHs in these halos produce comparable masses by $z\sim10$.}
    \label{fig:Zoom_BH}
\end{figure*}

In Fig. \ref{fig:Zoom_BH}, we show the evolution of the BH mass and luminosity of a single seed $\sim10^4~\rm{M}_\odot$ in the most massive halo in the Medium Halo ICs, again for the free-fall and Bondi accretion models under different feedback variations.
In this regime, the fiducial stellar and AGN feedback scenario~(top left panels) is extremely inhibitive to BH growth for all the accretion models. 
The accretion rates are well below Eddington~($f_{\rm edd}\sim0.01$) up to $z\sim10$, and only reach $f_{\rm edd}\sim0.1$ by $z\sim6$. 
Therefore, none of the accretion models produce any appreciable BH growth by $z=6$, with final BH masses not even exceeding $\sim2$ times the initial seed mass.

Interestingly, the impact of removing AGN feedback~(bottom left panels of Fig. \ref{fig:Zoom_BH}) is sharply distinct from what we found for $\sim10^5~M_{\odot}$ seeds in the Large-Halo ICs.
First, removing AGN feedback no longer produces any significant boost to BH growth in the Bondi model, increasing the final BH mass by only $\sim5\%$ relative to the fiducial AGN feedback scenario. 
Second, removing AGN feedback produces the strongest boost in the ff accretion model, where the accretion rate has the shallowest scaling with BH mass. 
The ff, modff, and Bondi models produce $z=6$ BH masses of $\sim7.46\times10^4~\rm{M}_\odot$, $\sim3.83\times10^4~\rm{M}_\odot$, and $\sim2\times10^4~\rm{M}_\odot$, respectively.
Therefore, in this regime, the seed mass is small enough that a steeper scaling between accretion rate and BH mass is less advantageous for BH growth. 
This contrasts with the $\sim10^5~M_{\odot}$ seeds in the Large-Halo ICs, where the model with the steepest scaling~(Bondi) produced the strongest BH growth when AGN feedback was removed.

However, the overall BH growth remains marginal across all accretion models even in the absence of AGN feedback, with none able to grow the BH by even a factor of $\sim10$ relative to its initial seed mass. 
This is primarily due to stellar feedback; reducing the strength of stellar feedback~(right panels of Fig.~\ref{fig:Zoom_BH}) yields a significant enhancement of BH growth for all accretion models, particularly at $z\lesssim10$. 
In fact, once all the models begin growing at the Eddington rate at $z\lesssim10$, they follow extremely similar BH growth tracks.
However, even with this sustained Eddington-limited growth, the onset occurs too late and the initial seed mass is too small to reproduce any of the JWST-inferred BH masses shown in Fig.~\ref{fig:Constrained_BH}, and we therefore do not show these AGN here.

\subsubsection{Individual $\sim10^3~M_{\odot}$ seeds in Small Halo ICs}\label{sec:Small}

\begin{figure*}
    \centering
    \includegraphics[width=1.0\linewidth]{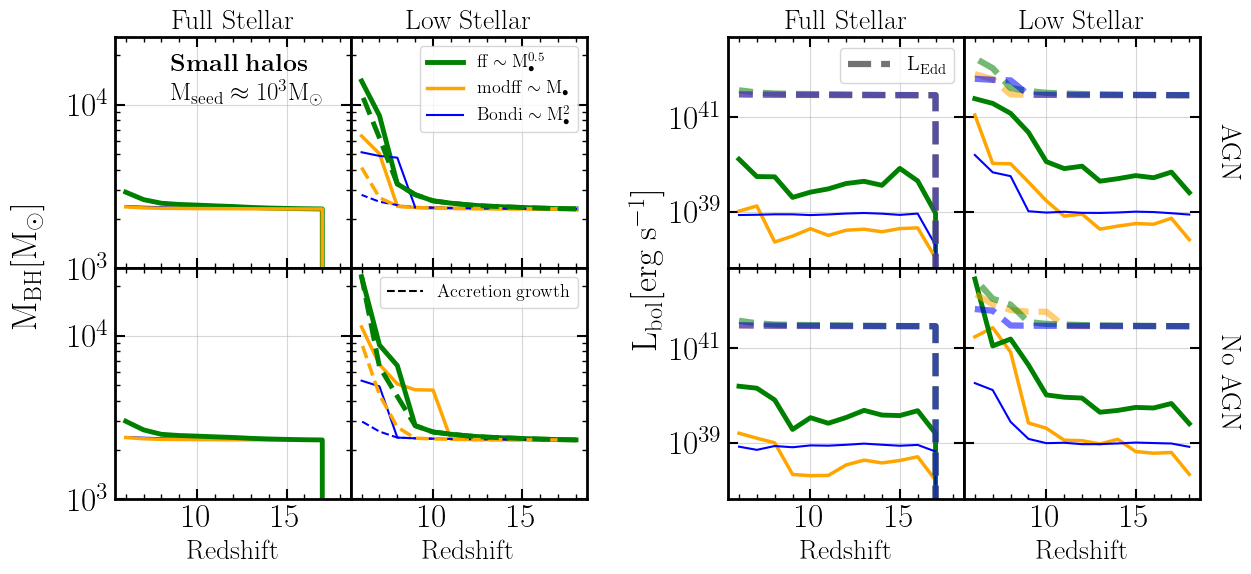}
    \caption{BH growth histories for the most massive BH in our small halo IC simulations with the strict seed model, again formatted the same way as in Fig. \ref{fig:Constrained_BH}. 
    In the right panels on the left, we show the accretion-driven growth as dashed lines.
    As in the medium halo mass case, we see negligible growth in the presence of fiducial stellar feedback.
    However, when stellar feedback is reduced here, BH growth is enhanced, but remains sub-Eddington.
    There is a slight enhancement in merger-driven growth in the low stellar feedback scenario, but this additional growth does not substantially change the resulting accretion-driven growth.}
    \label{fig:Small_BH}
\end{figure*}

Finally, in Fig.~\ref{fig:Small_BH}, we examine the growth of an even smaller $\sim10^3~M_{\odot}$ seed in the most massive halo in our Small-Halo ICs.
Here, we find that there is negligible BH growth for all three accretion models, even in the absence of AGN feedback. 
Consequently, AGN feedback has virtually no impact on BH growth in any of the models. 
Although the accretion rates and luminosities are higher for the ff model compared to Bondi or modff, the overall values~($f_{\rm edd}\sim0.01$--$0.1$) are so small that they have little effect on the BH mass.

As in the Medium-Halo ICs, it is only after reducing stellar feedback~(right panels of Fig.~\ref{fig:Small_BH}) that we begin to see substantial BH growth across all three models. 
Once again, the ff model produces the most growth, followed by modff and Bondi, indicating that models with steeper scalings between accretion rate and BH mass exhibit slower growth in this regime.

Notably, when stellar feedback is reduced, BH--BH mergers also contribute to the overall growth~(solid vs.\ dashed lines).
This is because reducing stellar feedback enhances star formation, leading to the formation of more seeds that can subsequently merge.
Here, dashed curves indicate the cumulative mass growth from accretion alone.
We find that in the Bondi model, which produces the least accretion, nearly all BH growth is driven by mergers. 
In contrast, for the ff and modff models, the majority of the mass accumulated by $z=6$ arises from accretion-driven growth, which begins at $z\sim9$.
Overall, we find that the only simulations that produce substantial accretion-driven growth of $\sim10^3~M_{\odot}$ seeds in a typical high-$z$ environment are those with the ff accretion model that has the shallowest scaling between accretion rate and BH mass, combined with a substantial reduction in stellar feedback.

\subsection{Lenient Black Hole Seed Models and the role of Mergers}\label{sec:lenient}

\begin{figure}
    \centering
    \includegraphics[width=0.9\linewidth]{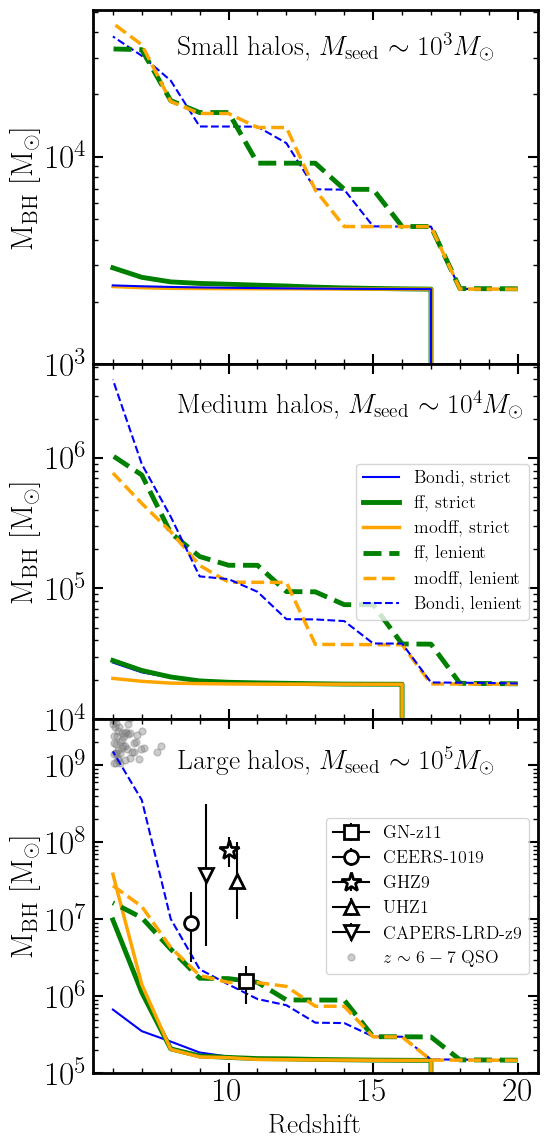}
    \caption{BH growth histories for the most massive BH in our three different ICs with strict (solid lines) and lenient (dotted lines) seeding criteria. 
    The fiducial AGN and stellar feedback was applied for all of the runs presented in this figure. 
    More lenient seeding allows for enhanced BH growth from BH--BH mergers at early times, which preferentially enhances Bondi growth for $10^4~\&~10^5~M_{\odot}$ seeds in Medium and Large Halo ICs respectively. 
    This is due to the stronger ($M_{\bullet}^2$) scaling of Bondi accretion with BH mass than ff or modff.}
    \label{fig:lenient}
\end{figure}

We have thus far used strict seed models to compare the ability of different accretion models to grow single BH seeds, and also demonstrated how early BH growth is strongly suppressed by our fiducial assumptions of stellar and AGN feedback regardless of the accretion model.
In this section, we keep our fiducial AGN and stellar feedback assumptions, but switch to our lenient seed models to understand how that impacts our BH growth under different accretion models.
In Fig. \ref{fig:lenient}, we show the resulting evolution of the most massive BH in all three ICs and their respective seed masses~(dashed lines), while also comparing against our strict seed model results~(solid lines).  

Across all three ICs, the immediate consequence of switching to a lenient seed model is a boost in early BH growth coming from BH--BH mergers. 
Due to the suppression of early accretion by stellar and AGN feedback, these mergers essentially dominate these earliest phases of BH growth. 
This has been demonstrated in the \texttt{BRAHMA} simulations~\citep{bhowmick2025dynamicslowmassblackhole,Bhowmick_etal_2026} with the Bondi accretion model, but here we show that this is also true under the free-fall models. 
In fact, in the Small Halo ICs~(top panel of Fig. \ref{fig:lenient}), there is no significant difference in the BH mass growth histories for any of the accretion models, since the growth is entirely merger-dominated all the way to $z=6$.

For the Medium Halo and Large Halo ICs~(middle and bottom panels of Fig. \ref{fig:lenient}), BH growth is similar between the three models up until $z\sim9$, as the growth is dominated by mergers. 
At $z\lesssim9$, however, as accretion-driven growth begins to pick up for all three accretion models, the Bondi case starts to grow faster than the ff and modff models. 
This is a direct consequence of the steeper $M_{\rm \bullet}^2$ scaling of the Bondi accretion rate. 
Due to their shallower scalings, the free-fall accretion models cannot take sufficient advantage of the initial boost in BH mass provided by mergers.

The differences in the behavior of our three accretion models have important implications for observed high-$z$ BH populations. 
First, the lenient seed models produce sufficient mergers in the Large Halo ICs to reach $\sim2\times10^6~M_{\odot}$ by $z\sim10.5$, consistent with current estimates for GN-z11. 
By $z\sim8.5$ (corresponding to the observed redshift of CEERS-1019), accretion-driven growth begins to contribute significantly in the Bondi model, reaching $\sim5\times10^6~M_{\odot}$, marginally consistent with current error bars on the measurement of CEERS-1019. 
The free-fall models exhibit weaker accretion-driven growth and fall slightly below the inferred BH mass for CEERS-1019. 
The most striking differences, however, arise for the brightest $z\sim6$ quasars powered by $\gtrsim10^9~M_{\odot}$ BHs. 
Such masses are achieved only in the Large Halo ICs with the Bondi accretion model, owing to runaway growth triggered by early BH--BH mergers. 
In contrast, both free-fall models produce only $\sim10^7~M_{\odot}$ BHs by $z\sim6$, nearly two orders of magnitude below the masses inferred for observed quasars.
For the Bondi accretion model, this represents a two-phase assembly pathway for SMBHs: BH growth at high-$z$ is driven primarily by many BH--BH mergers, enabling runaway accretion-driven growth at relatively later times. 
This has also recently been demonstrated in the much larger $[369\,\mathrm{Mpc}]^3$ \texttt{AMBRA} cosmological simulation~\citep{2026arXiv260401123Z}, which adopted a lenient version of the \texttt{BRAHMA} seed model coupled with Bondi accretion through the underlying \texttt{ASTRID} galaxy-formation model~\citep{2022MNRAS.512.3703B,2022MNRAS.513..670N}.

\section{Discussion}\label{sec:Discussion}

The main objective of this work was to compare the impact of the Bondi and free-fall accretion models on high-$z$ BH growth. 
We find that none of the models universally produces stronger early BH growth than the others we tested.\footnote{Note that the ff and modff models were tuned to the z=0 BHMD, and could likely reproduce the required high-z growth with an alternative tuning of the model parameters.} 
Instead, their relative behavior depends strongly on the initial seed mass, seed abundance, host environment, and the strengths of stellar and AGN feedback. 
This is determined by the complex interplay of trends that arise from two key differences between the models: (i) the steeper mass scaling of the Bondi accretion rate produces a stronger tendency toward runaway BH growth compared to the free-fall models, while (ii) the sound-speed dependence of the Bondi model makes it more susceptible to temperature increases from AGN and stellar feedback.

\subsection{The Impact of BH Mass Dependence on Accretion Rates}

A steeper scaling between accretion rate and BH mass naturally makes it easier for higher-mass BHs to grow, at the expense of lower-mass BHs. 
We see this trend most clearly in simulations without AGN feedback. 
For higher-mass $\sim10^5~M_{\odot}$ seeds in the Large Halo ICs, the Bondi model has a clear advantage over the free-fall models, but the trend is fully reversed for $\sim10^3~M_{\odot}$ seeds in the Small Halo ICs (see bottom-right panels on left of Figs. \ref{fig:Constrained_BH}, \ref{fig:Small_BH}). 
Even among the free-fall models, the modified freefall model~($\dot{M}_{\rm \bullet}\propto M_{\bullet}$) produces stronger growth than the FF model for $\sim10^5~M_{\odot}$ seeds in the Large Halo ICs, while the reverse is true for $\sim10^3~M_{\odot}$ seeds in the Small Halo ICs.

Additionally, when early BH growth is boosted by an abundance of BH--BH mergers, accretion models with stronger BH-mass scalings (i.e. Bondi) produce enhanced accretion-driven growth at later times. 
This effect naturally becomes more pronounced with increasing seed mass, as it strongly manifests for $\sim10^4~\&~10^5~M_{\odot}$ seeds in the Medium and Large Halo ICs, but not for $\sim10^3~M_{\odot}$ seeds in the Small Halo ICs.

\subsection{The Impact of Gas Densities and Temperatures: Role of Stellar and AGN Feedback} \label{sec: density and temperature}

\begin{figure*}
    \centering
    \includegraphics[width=1.0\linewidth]{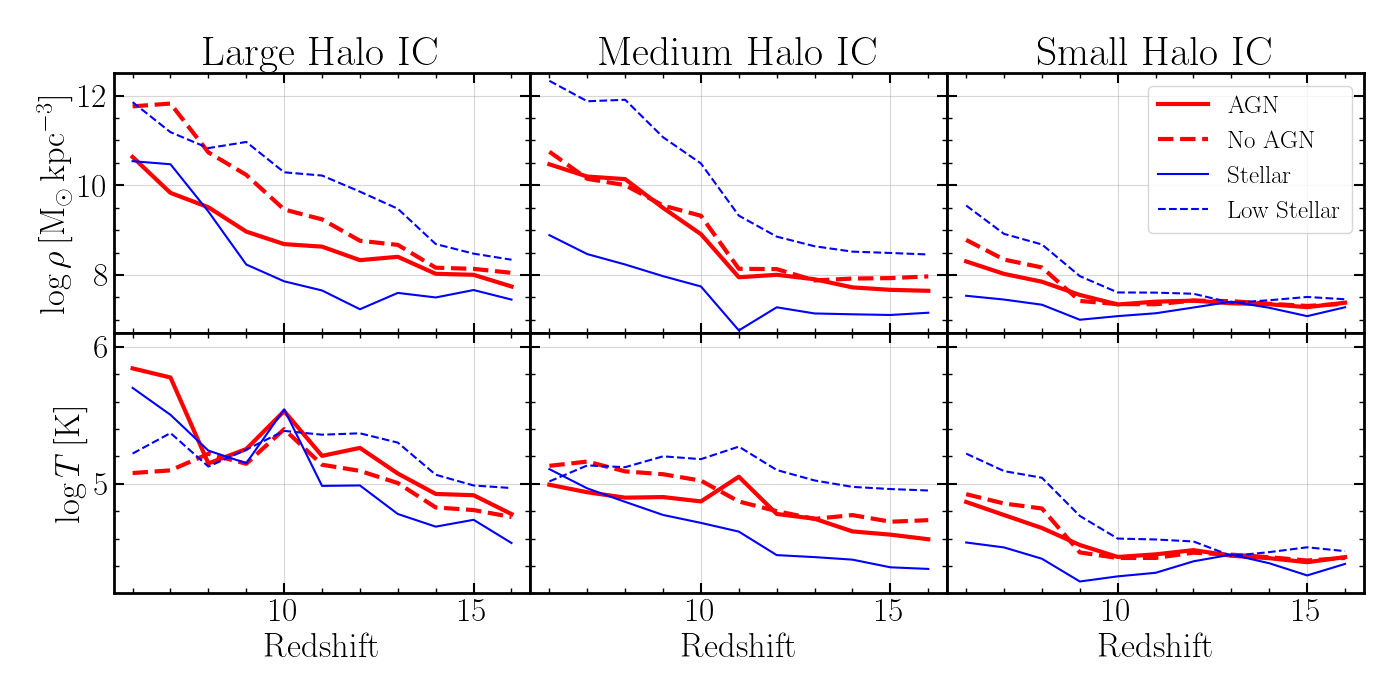}
    \caption{Redshift evolution of the average gas temperature and density in the most massive BH's kernel, averaged over all three accretion models for each IC.
    For each line plotted, we also average over variations in the non-specified feedback channel, i.e. the `AGN' line averages over both AGN and weak feedback runs as well as AGN and fiducial feedback runs.
    The large increase in temperature at late times for the large halo IC explains the minimal BH growth that the Bondi model exhibits in the fiducial AGN and stellar feedback regime. 
    The very large densities in the medium halo ICs with reduced stellar feedback also explains the consistent Eddington-limited accretion growth of the BHs in this IC regardless of accretion model.}
    \label{fig:T_rho_z}
\end{figure*}

When applied, the fiducial stellar and AGN feedback readily overshadow the trends one might intuitively expect from the BH mass scaling, due to their effects on the gas density and temperature. 
In the Small as well as Medium Halo ICs, the impact of stellar feedback is so strong that BHs undergo very little growth regardless of which accretion model is used. 
This is primarily because of the impact of stellar feedback on decreasing the overall gas densities, as shown in Fig. \ref{fig:T_rho_z}~(upper middle and right panels). 
The impact of AGN feedback on the densities and temperatures is also much smaller in the Medium and Small Halo ICs since the resulting accretion rates are so low.

In the Large Halo ICs, the impact of AGN feedback on the gas densities and temperatures is much stronger, and actually becomes larger than stellar feedback by $z\lesssim9$~(compare left and middle panels of Fig. \ref{fig:T_rho_z}). 
While both Bondi and free-fall accretion is suppressed by the reduction of gas density, Bondi is additionally penalized by the increase in gas temperature, as it also depends inversely on the local gas sound speed. 
The resulting suppression of Bondi accretion is so strong that it loses all the advantage provided by the high seed mass and the $M_{\bullet}^2$ scaling, and its final $z=6$ BH mass falls below that of the free-fall accretion models.

{}

\subsection{Implications for the observed high-z BHs}\label{sec:implications}

Among our Large Halo IC simulations with fiducial assumptions for AGN and stellar feedback, the only ones that produce the $\sim10^9~M_{\odot}$ BHs powering high-$z$ quasars are those in which Bondi accretion is combined with our lenient seed models. 
This is because the early boost from merger-driven BH growth initiates subsequent runaway accretion due to the $M_{\rm \bullet}^2$ scaling of the Bondi model.
The free-fall accretion models simply do not benefit from these BH--BH mergers to the same extent because of their weaker BH-mass scalings. 
Therefore, even if the free-fall models may be more physically justified for cold gas accretion, the existence of high-$z$ quasars suggests that an additional Bondi-like accretion mode may be needed, which can be efficiently amplified by the merger-driven growth produced in case of abundant formation of heavy seeds.

Among the $z\sim9-11$ JWST AGNs, only the BH mass inferred for GN-z11~($1.58\times 10^{6}\rm{M}_\odot$ at $z=10.6$) can be reproduced through abundant heavy seeds and BH--BH mergers under fiducial assumptions for AGN and stellar feedback.\footnote{This comparison is very approximate, as an accurate comparison would require reliable number densities for GNz-11-like systems, which are currently lacking in the literature.}
This is true regardless of whether we adopt the Bondi or free-fall accretion model, since the growth is dominated by BH--BH mergers. 
In the presence of such abundant mergers, all the accretion models can also marginally produce BH masses close to that inferred for CEERS-1019~($8.9\times10^6\rm{M}_\odot$ at $z=8.7$), although they lie closer to the lower end of the current error estimate. 
The estimates for objects such as UHZ1, GHZ9 and CAPERS-LRD-z9~($\sim10^7-10^8~M_{\odot}$) cannot be produced under fiducial stellar and AGN feedback assumptions. 
As shown by \cite{Bhowmick_etal_2026}, this is basically because the mergers are neither rapid nor efficient enough. 
However, it is worth noting that both these BH mass estimates---and even their AGN interpretation---are currently being actively debated~\citep{Naidu_etal_2025, Rusakov_etal_2026, Zou_etal_2026}. 
In any case, our current tentative conclusion is that, under fiducial stellar and AGN feedback assumptions, reproducing the currently inferred $z\sim9-11$ JWST BH masses is not viable in our simulations regardless of whether we adopt the Bondi or free-fall accretion model. 
Instead, abundant heavy seed formation and the resulting BH--BH mergers are necessary to boost early BH growth and produce some of the JWST BHs.

At the same time, there is also no compelling reason to rule out the possibility that the fiducial stellar and AGN feedback models required to reproduce low-$z$ galaxies and BHs remain valid at high redshifts. 
In fact, the JWST discovery of UV-luminous high-$z$ galaxies~\citep{Harikane_etal_2023, castellano_early_2023, finkelstein_complete_2024} has proven challenging to reproduce in many simulations~\citep{kannan_millenniumtng_2023, yung_are_2024, fujimoto_uncover_2024, lu_galaxy_2026}. 
One of the possible solutions currently being explored is weaker stellar feedback, or effectively feedback-free starbursts, at early times~\citep{dekel_efficient_2023, libanore_effects_2024, li_feedback-free_2024, somerville_density-modulated_2025, fujimoto_primordial_2025}. 
Additionally, there are several physical scenarios under which AGN feedback may be substantially weaker than for our fiducial assumptions. 
For example, in the slim-disk model of super-Eddington accretion, the accretion disk puffs up into a geometrically thick torus, increasing the photon diffusion timescale~\citep{Abramowicz_etal_1988, Sadowski_Narayan_2016}. 
This may effectively translate to a reduction in the radiative efficiency and therefore the feedback energy coupled to the surrounding gas\footnote{We acknowledge that super-Eddington accreting BHs may still be able to produce strong jets if they are rapidly spinning \citep{curd_grrmhd_2019}. However, while many accreting SMBHs in the local Universe appear to be spinning rapidly \citep{reynolds_observational_2021, sisk-reynes_spin_2026}, it is unclear if this extends to SMBHs in the very early Universe or high Eddington-ratio accreting BHs.}. 
Furthermore, in dense and clumpy high-redshift environments, AGN radiation may preferentially escape through low-density channels without coupling efficiently to the gas~\citep{Ward_etal_2024}.

\subsection{Modeling caveats}

Our main findings regarding the relative performance of the Bondi and free-fall models may also be substantially influenced by features of our underlying galaxy formation model. 
First, the effective equation of state treatment of the ISM has a direct impact on the BH environmental properties used to determine the accretion rates. 
Due to the artificial pressurization of star-forming gas, the BHs experience an overly smoothed ISM that under-resolves dense gas structures. 
This can lead to an underestimation of the accretion rates, particularly for the free-fall models, which are more physically motivated for modeling cold gas accretion than the Bondi model. 
With that being said, the strong suppression of BH accretion by feedback has also been observed in many ``resolved ISM'' simulations  \citep{petersson_noctua_2025, Wellons_etal_2023}. 
Nevertheless, we cannot rule out the possibility that the free-fall accretion model may still reproduce the observed high-$z$ BH populations, and it would be instructive to explore this further using resolved ISM simulations in the future.

Additionally, the suppression of BH accretion by AGN feedback may depend sensitively on the specific feedback implementation adopted in \texttt{TNG}. 
In the current model, thermal energy injected isotropically within a spherical kernel can efficiently disrupt gas inflow from all directions. 
In contrast, if the AGN feedback were modeled as bipolar kinetic outflows, its impact on accretion along the orthogonal plane could be substantially reduced~(Partmann et al. in prep). 
Exploring such alternative feedback implementations will be an important direction for future work.

Finally, in all of our simulations we limited BH accretion rates to the Eddington rate. 
While BHs may in principle accrete at super-Eddington rates, the main conclusions of this work are unlikely to be significantly affected by allowing such growth. 
This is particularly true for our fiducial AGN and stellar feedback simulations, in which accretion rates remain well below the Eddington limit throughout most of the BH growth histories. 
The only cases in which super-Eddington accretion could substantially boost BH growth are the FreeFall models at $z\gtrsim9$ within Large-Halo ICs, where our simulations predict near-Eddington accretion~(revisit upper-left panels of Fig. \ref{fig:Constrained_BH}). 
While this may allow the FreeFall models to more readily reproduce the masses of the observed $z\sim6$ quasars, it is unlikely to enable the formation of the JWST-inferred $z\sim9$--11 BHs from individual $\sim10^5,M_\odot$ seeds, since the growth in this regime remains strongly sub-Eddington.

For simulations in which AGN or stellar feedback is reduced relative to the fiducial models, near-Eddington growth occurs across a broader range of accretion prescriptions and redshifts, wherein super-Eddington accretion could further enhance BH growth. 
However, in the absence of strong feedback regulation, the inclusion of super-Eddington accretion would primarily increase the overall normalization of BH growth, while being unlikely to alter the relative performance of the different accretion models.

\section{Conclusions}\label{sec:Conclusions}

In this paper, we analyze the impact of three different accretion models on early BH growth using a suite of cosmological simulations. Specifically, we compare the commonly used Bondi-Hoyle accretion formalism, in which the BH accretion rate scales as $M_{\rm \bullet}^2/c_s^3$, against two variants of the free-fall accretion model~(ff and modff), in which the accretion rate scales as $M_{\bullet}^{1/2}$ and $M_{\bullet}$, respectively. We explore these accretion models under different variations of the stellar and AGN feedback parameters in the \texttt{TNG} model, coupled with new BH seeding models developed as part of the \texttt{BRAHMA} simulation suite.

These accretion models have different dependencies on BH mass and the surrounding gas environment~(gas temperature and density).
The relative performance of the Bondi and free-fall models is therefore strongly influenced by the BH seed model, AGN and stellar feedback prescriptions, and the large-scale environment. More specifically, 

\begin{itemize}
\item For a single $\sim10^5~M_{\odot}$ seed within an extreme, overdense environment~($\sim10^{12}~M_{\odot}$ halo at $z=6$), Bondi accretion tends to outpace the free-fall accretion models when AGN feedback is absent. 
This is primarily due to the steeper $M_{\bullet}^2$ scaling of Bondi accretion, which leads to stronger runaway BH growth. 
However, once we include the AGN thermal feedback calibrated in \texttt{TNG} to reproduce low-$z$ BH observations, the temperature-sensitive Bondi accretion becomes disproportionately suppressed compared to the free-fall models, producing only a $\sim10^6~M_{\odot}$ BH by $z=6$. 
As a result, the free-fall models produce more massive $\sim10^7~M_{\odot}$ BHs by $z=6$ in the presence of AGN feedback, since their accretion rates do not explicitly depend on gas temperature.

\item The impact of AGN feedback is much weaker in less massive halos~($\sim10^{10}~M_{\odot}$ halo at $z=6$). 
For a single, lower-mass $\sim10^3~M_{\odot}$ seed growing in such a region, the free-fall accretion models outpace Bondi due to their shallower $M_{\bullet}^{1/2}$ and $M_{\bullet}$ scalings, regardless of whether AGN feedback is included. 
However, stellar feedback becomes a strong regulator of BH growth in these halos. 
When we apply the level of stellar feedback required to reproduce the observed galaxy stellar mass functions in \texttt{TNG}, we find that BH growth up to $z=6$ is almost negligible compared to the original seed mass.

\item If we adopt more lenient seeding models and form larger numbers of seeds, BH growth is boosted by mergers in regimes where accretion is otherwise suppressed by feedback. 
In overdense environments, this leads to a disproportionately stronger enhancement of accretion in the Bondi model compared to the free-fall models, due to the runaway growth driven by the steeper scaling with BH mass.

\end{itemize}

Under the standard \texttt{TNG} model's AGN feedback assumptions, our simulations cannot produce any of the current mass estimates of JWST-discovered BHs at $z\sim9-11$, regardless of whether we use Bondi or free-fall accretion. 
If AGN feedback indeed regulates BH accretion as strongly as in our simulations, abundant heavy $\sim10^5~M_{\odot}$ seeds can still boost early BH growth through BH--BH mergers and assemble BHs up to $\sim10^6-10^7~M_{\odot}$ by $z\sim9-11$. 
This can explain the inferred masses of GN-z11 and CEERS-1019, but the existence of $\gtrsim10^7~M_{\odot}$ BHs at $z\gtrsim9$ would still pose a significant challenge. 
Finally, assembling $\sim10^9~M_{\odot}$ BHs by $z\sim6$~(consistent with the brightest quasars known prior to JWST) under these AGN feedback assumptions requires not only abundant mergers, but also subsequent runaway BH growth that is only possible in our simulations with Bondi accretion.

If the formation of heavy seeds in such abundant numbers is not possible, another avenue for boosting early BH growth is reduced AGN feedback at early times combined with Bondi accretion-driven runaway growth. 
In such a scenario, however, we would still need to understand how AGN feedback later strengthens to prevent the overgrowth of BHs at lower redshifts.

Overall, Bondi appears to have an advantage over the free-fall models in accelerating early BH growth, particularly for heavy seeds in rare extreme environments, due to its stronger tendency for runaway growth. 
However, our effective equation of state-based ISM treatment is also a disadvantage for the free-fall accretion models, since the cold dense phases of the ISM are not explicitly resolved.
Therefore, future work will continue exploring these free-fall models in simulations that explicitly resolve the multiphase ISM.

\section{Acknowledgements}

Our work is part of the ``Learning the Universe'' collaboration, which aims to understand the extragalactic Universe by jointly inferring the initial conditions and physical laws that govern its subsequent evolution. AKB and PT acknowledge support from NSF-AST 2510738.
JK, AKB, PT, and AMG acknowledge support from NSF-AST 2346977 and the NSF-Simons AI Institute for Cosmic Origins which is supported by the National Science Foundation under Cooperative Agreement 2421782 and the Simons Foundation award MPS-AI-00010515.
RW acknowledges funding of a Leibniz Junior Research Group (project number J131/2022).
The authors acknowledge Research Computing at The University of Virginia for providing computational resources and technical support that have contributed to the results reported within this publication. URL: \href{https://rc.virginia.edu}{https://rc.virginia.edu}.

\bibliography{biblio}{}
\bibliographystyle{aasjournalv7}

\end{document}